# Cation Engineering of Cu-Doped CsPbI₃: Lead Substitution and Dimensional Reduction for Improved Scintillation Performance

*David Hadid Sidiq,‡ Somnath Mahato,*,‡ Tobias Haposan,‡,♣ Michal Makowski,‡ Dominik Kowal,‡ Marcin Eugeniusz Witkowski,ᵀ Winicjusz Drozdowski,ᵀ Arramel,‡,* and Muhammad Danang Birowosuto‡,\**

‡Center of Excellence Applied Physics and Chemistry, Nano Center Indonesia, Jalan Raya PUSPIPTEK, South Tangerang 15314, Indonesia

♣Lukasiewicz Research Network - PORT Polish Center for Technology Development, Stabłowicka 147, Wroclaw 54-066, Poland

♣Department of Materials Science and Engineering, National University of Singapore, 9 Engineering Drive 1, Singapore 117575, Singapore

ᵀInstitute of Physics, Faculty of Physics, Astronomy, and Informatics, Nicolaus Copernicus University in Torun, ul. Grudziadzka 5, 87-100 Torun, Poland

**KEYWORDS**: Cesium lead iodide, copper-doped, decay time, radioluminescence and scintillation.

## ABSTRACT

To date, inorganic halide perovskite nanocrystals show promising contributions in emerging luminescent materials due to their high tolerance to defects. In particular, the development of cesium lead iodide (CsPbI₃) has shown its efficiency for light-harvesting properties. However, further implementation is hindered due to the toxicity of the lead content. Therefore, in this study, we introduced Cu atoms to partially substitute Pb atoms (5% Cu) in the CsPbI₃ lattice as a solution to reduce Pb toxicity. A partial lead material is substituted using Cu displays a larger Stokes shift (~67 nm) compared to the pristine, and resulted doped-CsPbI₃ not undergo the undesired self-absorption. An outcome is focused on the champion of fast-component ($\tau_1$) decay time ~0.6 ns. Temperature-dependent radioluminescence outlines an incremental change in the emission intensity is marginally centered at 713 ± 16 nm, which indicates Cu-doped CsPbI₃ is not greatly affected by temperature. In addition, we report that the light yield (LY) pristine CsPbI₃ after doping is increased to 3.0 ± 0.8 photons/keV. Our work provides physical insights into a tunable scintillation property using transition metal doping toward lead-free based scintillating perovskites.

## 1. Introduction

The development of scintillating materials with high efficiency and fast detection is a crucial area of research in the field of ionizing radiation detection. The successful application of halide perovskites for the direct detection of X-rays has significantly expanded the potential of halide perovskites in the detection of high-energy radiation, thereby encouraging further research on their scintillation properties. Lead halide perovskites known mainly in the form of $CsPbX_3$ (where X is a halogen such as I, Br, or Cl), are emerging as a promising class of scintillating materials due to their distinctive combination of optical, electronic, and structural properties.[1] These optical properties include a high-light yield (LY), a fast decay time,[1] a narrow emission bandwidth (12-42 nm), and the ability to cover the entire range of the visible spectrum. The material exhibits a high quantum yield ($\leq 90\%$), a radiative lifetime in the range of 1-29 ns,[2] a high-light absorption coefficient, tunable optical emission, an affordable cost, high defect tolerance,[3] a strong emission intensity under high energy excitation, and a fast glow decay. Nevertheless, lead halide perovskites continues to exhibit certain limitations due to its requirement for low temperatures below 100 K for emission.[4,5] Lead halide perovskites display a relatively low LY (less than 1 photon/kV)[6,7] and exhibit sensitivity to temperature, water, light, and heat at room temperature.[8]

Among the lead halide perovskites, cesium lead iodide ($CsPbI_3$) has been identified as a promising candidate due to its enhanced stability relative to other lead halides (Cl and Br).[9] Additionally, this material exhibits an extended visible light absorption tail compared to $CsPbBr_3$ and $CsPbCl_3$.[10] The average decay time of $CsPbI_3$ has been reported to be 1.10 ns faster than $CsPbBr_3$ (5.97 ns) and therefore highly suitable as a fast-detecting scintillator.[11] However, its low stability at room temperature necessitates further development.[8,10,12–15] Additionally, $CsPbI_3$ exhibits a small Stokes shift, which causes self-absorption and renders its function is unsuitable for a scintillator material.[11] A variety of techniques have been employed to enhance the scintillating properties of $CsPbI_3$, including the introduction of metal ions and the reduction of its dimension.[8] The incorporation of metal ions has the potential to enhance the scintillating ability due to its defect-tolerant ionic lattice.[13] Several metal cations is used to enhance the optical characteristics of CsPbI₃, including $Sr^{2+}$, $Cu^{2+}$, $Co^{2+}$, $Zn^{2+}$, $Mn^{2+}$, $Ag^{2+}$, and $Ni^{2+}$.[8,10,12–15] The generic approach is to incorporate the substituted metal ions with a smaller atomic radius. In general, the introduction of metal doping can influence the stability, phase, and optoelectronic properties of the material. Nevertheless, CsPbI₃ applications in scintillator enhancement have yet to be widely documented, particularly concerning the holy grail of LY enhancement.

In terms of crystal structure, CsPbI₃ has a distinct cubic structure (α-phase) exhibits an excellent optical properties.[10] However, α-(cubic) can only maintain its phase at temperatures above ~230°C and undergoes a phase conversion to a non-perovskite orthorhombic structure (δ-phase) when the temperature is below 230°C.[16] Meanwhile, the orthorhombic structure (γ-phase) is stable at low temperatures (<10°K) and structurally unstable at room temperature. We also note that the tetragonal structure (β-phase) exhibits a superior stability compared to the orthorhombic structure (γ-phase), rendering its promising candidate for optoelectronic devices.[12] The incorporation of Cu into CsPbI₃ crystals has been observed to induce alterations in the initial phase of CsPbI₃, reduce the lattice size, enhance the lifetime decay, and elevate the PLQY.[3,8,10] These expected outcomes are aligned with the conventional requirements for scintillator applications. In this paper, we investigate the impact of Cu towards dopant engineering on the host lattice of lead halide perovskite at a slight low concentration of 5%, exhibiting unprecedented tetragonal formation. Furthermore, the impact of dimensional reduction is examined in conjunction with Cu doping to ascertain the influence Cu atomic substitution on optical properties for scintillator applications.

## 2. Experimental section

### 2.1. Synthesis

The synthesis of CsPbI₃ nanocrystals according to the previous work.[10] First, cesium oleate (Cs-OA), 814 mg of calcium bicarbonate ($Cs_2CO_3$), and 2.5 mL of oleic acid (OA) were introduced into 40 mL of 1-octadecene (ODE) solvent. A subsequent heating at 120°C for 1 hour under an inert atmosphere was carried out. The temperature solution was increased to 150°C until the color became transparent, then it was gently cooled to room temperature. Second, 870 mg of lead(II) iodide ($PbI_2$) and 5 mL of ODE were mixed and simultaneously heated at 120°C for 1 hour. Subsequently, 1 mL each of OA and oleylamine (OLA) were swiftly injected and stirred under an inert atmosphere, and the temperature was raised to 150°C. During this period, 4 mL of freshly preheated (100°C) Cs-OA solution was injected when the solution turned transparent. The successful reaction when the solution is turned to red, implying $CsPbI_3$ nanocrystals was completed, then a reaction was terminated by cooling the product using an ice bath. To purify the resulting perovskite dispersion, an excess of hexane was added, followed by repeated centrifugation at 20,000 rpm to remove unreacted OA/OLA. The precipitate was collected and redispersed in hexane for further characterization. The synthesis of Cu-doped $CsPbI_3$ nanocrystals was followed by previous work with different ratios.[10] Anhydrous copper acetate ($Cu(CH_3COO)_2$) was used as the $Cu^{2+}$ ion source. Separately, $CsCu_2I_3$ and $Cs_3Cu_2I_5$ nanocrystals were synthesized mechanochemically using cesium iodide (CsI) and copper iodide (CuI) in molar ratios of 1:2 and 3:2, respectively, based on a previous method.[17] The mixtures were each ground for 10 minutes with 300 μL of acetone, resulting in the formation of $CsCu_2I_3$ and $Cs_3Cu_2I_5$ nanocrystals, respectively. All chemicals, including $Cs_2CO_3$, $PbI_2$, OA, OLA, hexane, ODE, CsI, and CuI, were obtained from Sigma-Aldrich and used without further purification.

2.2 Transmission electron microscopy

Transmission electron microscopy (TEM) and high-resolution TEM (HRTEM) analyses were undertaken utilizing a double-Cs-corrected Titan G2 60–300 (S)TEM microscope. Before HRTEM analysis, the electron beam conditions were adjusted based on the sample material. In this study, an acceleration voltage of 300 kV was employed. Samples were prepared by dissolving 4 mg of the compound in 4 mL of ethanol, followed by 1-hour sonication. The samples were transferred to Cu holey carbon grids (~5 uL droplet) and allowed to dry. Both TEM and STEM modes were employed, also at very high magnifications allowing high-resolution imaging. For image acquisition, the Ultrascan US1000 camera was used for TEM, and the HAADF detector for STEM. The STEM-energy dispersive X-ray spectroscopy (EDS) maps were also acquired with a Super-X EDS detector. Due to sample sensitivity to the beam, the data was collected with the smallest possible dose at the expense of signal-to-noise ratio, which is particularly seen on EDS maps.

2.3 Absorption-photoluminescence spectroscopy

The samples were excited for photoluminescence (PL) measurements using a picosecond laser diode (Master Oscillator Fibre Amplifier, Picoquant GmbH, Berlin, Germany) with a repetition rate of 31.25 kHz and a wavelength of 266 and 532 nm for copper and lead halide perovskite nanocrystals, respectively. The laser diode had a pulse duration of 50 ps and an average power of 2 mW at RT. Excitation focusing and signal collection were carried out using a microscopic objective (Nikon Corporation, Tokyo, Japan) with a magnification of 20x and a numerical aperture of 0.4. The PL signal was filtered and acquired using a high-sensitivity visible light spectrometer (Ocean Optics, Florida, USA). Absorption spectra were measured with a homemade setup with the same commercial spectrometer in transmission mode. Since the samples were put inside the quartz tubes, the spectra were corrected for the absorption of the tubes.

2.4 Time-resolved photoluminescence spectroscopy

Time-resolved photoluminescence spectroscopy (TR-PL) measurements are carried out with the repetition rate of 31.25 kHz, and the PL signal (selected by a longpass filter for wavelengths longer than 500 and 600 nm for copper and lead halide perovskite nanocrystals, respectively) was directed to a single-photon avalanche photodiode (APD). The timing response was analyzed using time-correlated single-photon counting electronics (HydraHarp 400, PicoQuant, Germany). Absorption data were gathered using a custom apparatus, integrating an Avaspec device in transmission, and corrections were made considering the quartz tube's absorption properties.

2.5 Temperature-dependent radioluminescence measurements

An Inel XRG3500 X-ray generator (Cu-anode tube, 45 kV/10 mA) was used as an X-ray source. An Acton Research Corporation SpectraPro-500i monochromator (500 nm blazed grating) coupled with a Hamamatsu R928 photomultiplier was used to record the RL signal. An APD Cryogenics Inc. closed-cycle helium cooler with a Lake Shore 330 programmable temperature controller was to control the temperature of the sample between 10 and 350 K. To avoid any glow effects from the thermal release of charge carriers, measurements were performed starting at 350 K and terminating at 10 K.

## 3. Results and Discussion

Figure 1a-d displays the optimized crystal visualization of respective all-inorganic perovskites discussed in this study.[18] The atomic arrangement of $CsPbI_3$ (Figure 1a) and Cu-doped $CsPbI_3$ (Figure 1b) are presented based on previous work by Roy and coworkers,[10] whereas the crystal structure of copper halide perovskite (CHP) $CsCu_2I_3$ (Figure 1c) and $Cs_3Cu_2I_5$ (Figure 1d) are collected from Haposan et al.[17] We confirmed defects within the crystal structure of $CsPbI_3$.[19] Figure 1b shows that Cu doping distorted the $CsPbI_3$ crystal structure. Theoretically, this defect did not initially introduce the undesired vacancies within the crystal. We further postulate here that the doping process also contributed to creating defects in the crystal structure of $CsPbI_3$.[11,20] Figure 1c and d show that the substitution of the Pb atom with Cu caused a crystal defect called a Frenkel defect.[17] The Frenkel defect creates vacancies in the crystal structure, which can affect the stability and detection time application of the scintillator. The XRD diffractogram of $CsPbI_3$ and Cu-doped $CsPbI_3$ with their corresponding Rietveld refinement are shown (Figure S1). We note that CIFs used for analysis are acquired from Roy et al.[10] and the extracted crystal lattice values are shown in Table 1.

The extracted crystal lattice parameters of $CsPbI_3$ were determined at 31.85 Å, 25.84 Å, and 18.84 Å for a, b, and c, respectively. In contrast, Cu doping attempt causing the a and b lattice parameters of pristine $CsPbI_3$ to decrease: 21.61 Å and 16.38 Å respectively. However, we note that the c lattice is increasing to 21.61 Å. Based on this geometric consideration, the atomic substitution of Pb with Cu atoms shows a reduction in the lattice sizes. In the case of $CsCu_2I_3$ compound, we note the a, b, and c lattice sizes are 10.34 Å, 13.12 Å, and 6.35 Å, respectively. This lattice size verifies that $CsCu_2I_3$ is considered as a one-dimensional (1D) material, as evidenced by the considerable difference in the dimensions of its c lattice compared to the other lattice parameters. On the other hand, $Cs_3Cu_2I_5$ showed a slight decrease in the b crystal lattice size from 13.12 Å to 11.77 Å, and underwent a significant increase in the c crystal lattice size from 6.35 Å to 14.57 Å. Based on the previous DFT results (Table 1) indicated that $Cs_3Cu_2I_5$ categorized as a zero-dimensional (0D) material.[17]

The synthesis of the pristine form of $CsPbI_3$ resulted in the formation of an orthorhombic structure, designated as the γ-phase. Subsequently, upon doping with 5% Cu, $CsPbI_3$ underwent a phase transition to a tetragonal structure (β-phase), accompanied by a reduction in the crystal lattice size. This result differs from previous studies that employed a ratio of 5.6% Cu, which resulted in the initial phase of $CsPbI_3$ being a cubic structure (α-phase) and an orthorhombic structure (γ-phase). The phase change accommodates the Cu doped $CsPbI_3$ could be attributed to the merging of phases.[10] The observed decrease in the crystal lattice can be ascribed to the Pb atoms substitution with smaller Cu atoms (the ionic radius of $Cu^{2+}$ and $Pb^{2+}$ is 72 pm and 119 pm, respectively).[8] Figure 1e presents a comparison of the absorption length of pristine $CsPbI_3$, Cu-doped $CsPbI_3$, $CsCu_2I_3$ (CHP 1D), and $Cs_3Cu_2I_5$ (CHP 0D). As demonstrated in Figure 1e, the incorporation of the dopant in the $CsPbI_3$ structure does not influence the absorption length significantly. Nevertheless, a reduction in the dimensionality order from $CsCu_2I_3$ to $Cs_3Cu_2I_5$ caused a decrease in the absorption length at the energies between 33 and 46 keV. We note that the overall Cu doping in the case of 3D $CsPbI_3$ resulted in a lower absorption length within the energy range between 1 to 1000 keV.

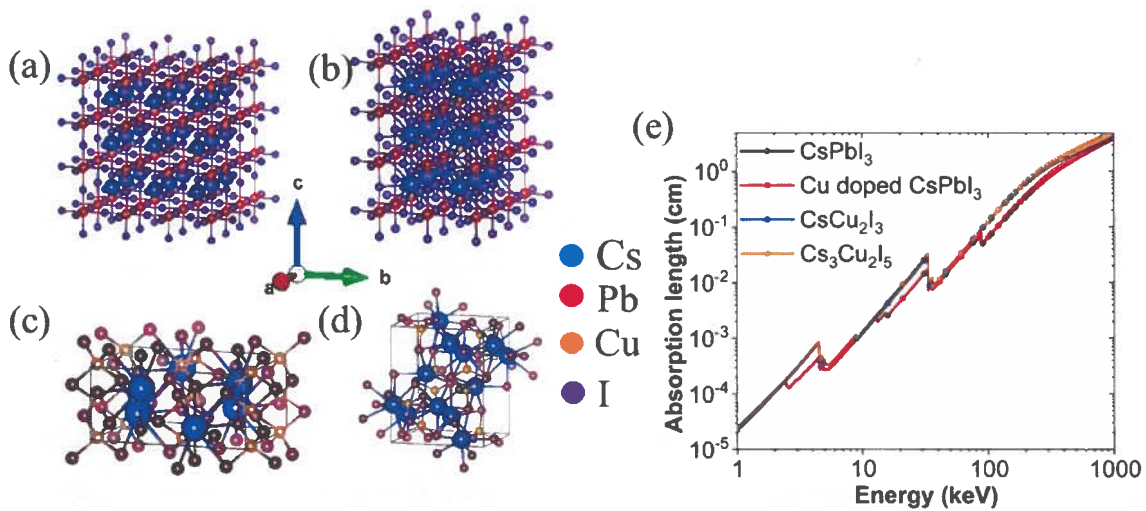

Figure 1. 3D structural visualization of the crystal structure of (a) $CsPbI_3$ (b) Cu-doped $CsPbI_3$ (c) $CsCu_2I_3$, and (d) $Cs_3Cu_2I_5$ using VESTA simulation, respectively. (e) X-ray absorption length for the investigated Cu effect for perovskite with $CsPbI_3$ was used. The crystal structures representations are plotted using VESTA.[18]

**Table 1.** Crystallographic parameters of CsPbI$_3$, Cu-doped CsPbI$_3$, CsCu$_2$I$_3$, and Cs$_3$Cu$_2$I$_5$.

| Compounds | CsPbI$_3$ | 5%Cu-CsPbI$_3$ | CsCu$_2$I$_3$ | Cs$_3$Cu$_2$I$_5$ |
|---|---|---|---|---|
| Space group | Pnma | P4/mbm | Cmcm | Pnma |
| a (Å) | 31.85 | 21.61 | 10.34 | 10.39 |
| b (Å) | 25.85 | 16.38 | 13.12 | 11.77 |
| c (Å) | 18.84 | 21.61 | 6.35 | 14.57 |
| Alpha (°) | 88.00 | 88.00 | 90.00 | 90.00 |
| Beta (°) | 88.00 | 88.00 | 89.99 | 90.00 |
| Gamma (°) | 90.00 | 90.00 | 89.98 | 90.00 |
| Volume (Å$^3$) | 15510.67 | 7648.26 | 860.69 | 1782.05 |
| R$_{wp}$ | 31.70 | 48.30 | 26.12 | 12.16 |
| $\chi^2$ | 9.68 | 9.18 | 15.29 | 9.51 |
| Ref. | 10 | 10 | 17 | 17 |

The morphological state of the pristine CsPbI$_3$ was confirmed using transmission electron microscopy (TEM), as illustrated in Figure 2a. The crystal size of the orthorhombic CsPbI$_3$ is observed to vary in length, with an irregular arrangement, and to range from approximately 15 to 17 nm. Upon introduction of Cu doping into the host lattices, demonstrates Cu doping alters the orthorhombic arrangement of the CsPbI$_3$ crystals, as illustrated in Figure 2b. The interplanar spacing of pristine and Cu-doped sample is shown in Figure 2d and e, respectively. Upon 5% Cu doping, the interplanar spacing is slightly reduced to 0.658 ± 0.0269 nm. This result lends support to the hypothesis of previous work, which was a reduction of 0.03 nm for 5.6% doping.[10] We note, 1D CsCu$_2$I$_3$ display a triangular shape (Figure 2c), while in the case of Cs$_3$Cu$_2$I$_5$ illustrates a circular shape (Figure 2f), indicating the arrangement of 0D material. To determine the chemical composition of Cu-doped $CsPbI_3$, we refer to the reported XPS measurements as a reference.[10]

The optical properties were analyzed by taking into account the absorption and photoluminescence (PL) spectra, as illustrated in Figure 3. The absorption peaks were identified based on the first excitonic peak. The maximum absorbances were observed at 591 nm, 612 nm, 318 nm, and 304 nm for $CsPbI_3$, Cu-doped $CsPbI_3$, $CsCu_2I_3$, and $Cs_2Cu_3I_5$, respectively. The maximum emission wavelengths were observed ~ 600 nm, 679 nm, 599 nm, and 482 nm for $CsPbI_3$, Cu-doped $CsPbI_3$, $CsCu_2I_3$, and $Cs_2Cu_3I_5$, respectively. In alignment with the conclusions of the $CsPbI_3$ preceding studies,[8–12,15,21,22] a cubic structure ($\alpha$-phase) is converted into an orthorhombic structure ($\gamma$-phase), yielded a red shift in the absorption peak upon Cu doping. Such optical response is a consequence of the B-site.[10] An analogous phenomenon related driven by the phase conversion from orthorhombic structure ($\gamma$-phase) to the tetragonal structure ($\beta$-phase) due to Cu doping is shown in this study. In the other hand, $CsPbI_3$ with the tetragonal structure ($\beta$-phase) exhibited a peak absorption, and PL emission was observed at the 736 nm and 716 nm, respectively.[12] This illustrates that $CsPbI_3$ with a tetragonal structure ($\beta$- phase) exhibited a blue shift in both the absorption and emission peaks when doped with Cu atoms, in comparison to the pristine $CsPbI_3$ ($\beta$-phase). The shift in peak absorption and PL emission due to the influence of Cu doping also increases the Stokes shift value from 9 to 67 nm. When compared to the $CsPbI_3$ ($\beta$-phase), the influence of Cu doping on $CsPbI_3$ is evidenced by a greater Stokes shift value.[12]

To compare the effect of dimensionality control, we present the optical properties of $CsCu_2I_3$ (1D) and $Cs_3Cu_2I_5$ (0D) since these compounds sharing similar criterion: All-inorganic and identical A-site and X-site[17,23–26] compared to the $CsPbI_3$ case, as depicted in Figure 3c and d, respectively. We note the impact of dimensional reduction resulted in a peak-shifting of both absorption peak and PL emission peak towards shorter wavelengths,[17] which eventually reduces the Stokes shift. The low-dimensional CHP family exhibits a larger Stokes shift compared to the three-dimensional lead halide perovskite.[11] The Stokes shift was observed to be 281 nm and 178 nm for $CsCu_2I_3$ and $Cs_3Cu_2I_5$, respectively.[17]

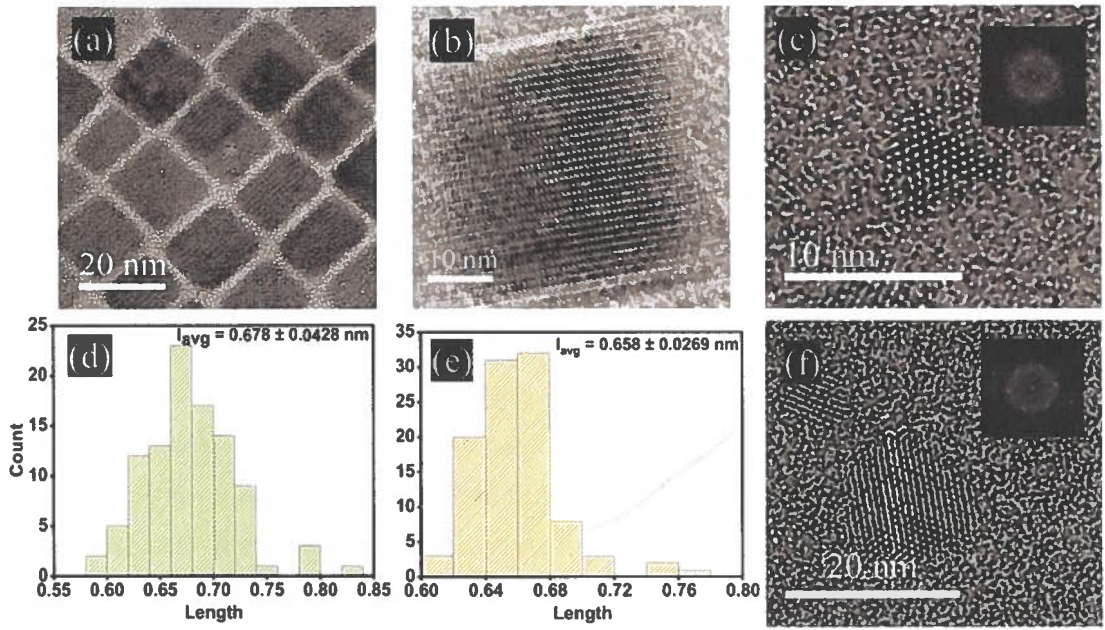

**Figure 2.** TEM images of (a) $CsPbI_3$ (b) Cu-doped $CsPbI_3$ (c) $CsCu_2I_3$, and (f) $Cs_3Cu_2I_5$. The TEM images for $CsPbI_3$ and Cu-doped $CsPbI_3$ equipped by average distribution interplanar space are shown in Figure (d) and (e) for $CsPbI_3$ and Cu-doped $CsPbI_3$, respectively. Cu-doped shows a decrease in interplanar space 2.98%. Figure (c) and (f) indicated that $CsCu_2I_3$ and $Cs_3Cu_2I_5$ formed in 1D and 0D respectively.

One of the primary objectives in the development of scintillators is the reduced of decay time, which accelerates the operational lifespan of a scintillator. Following the occurrence of radiation, three distinct components of decay are quantified. The fast component ($\tau_1$) associated with the direct electron hole capture process resulting a very fast charge carrier recombination, i.e. anomalous emission.[27] However, this component only appears in the scintillation decay curves of pristine and Cu-doped $CsPbI_3$ (Figure S2), but not in the case of $CsCu_2I_3$ and $Cs_3Cu_2I_5$.[11,17] The mid component ($\tau_2$) corresponds to the fast exciton contribution, and the slow phase ($\tau_3$) related with the electron-hole recombination due to a defect, i.e. self-trapping exciton (STE).[11,17]

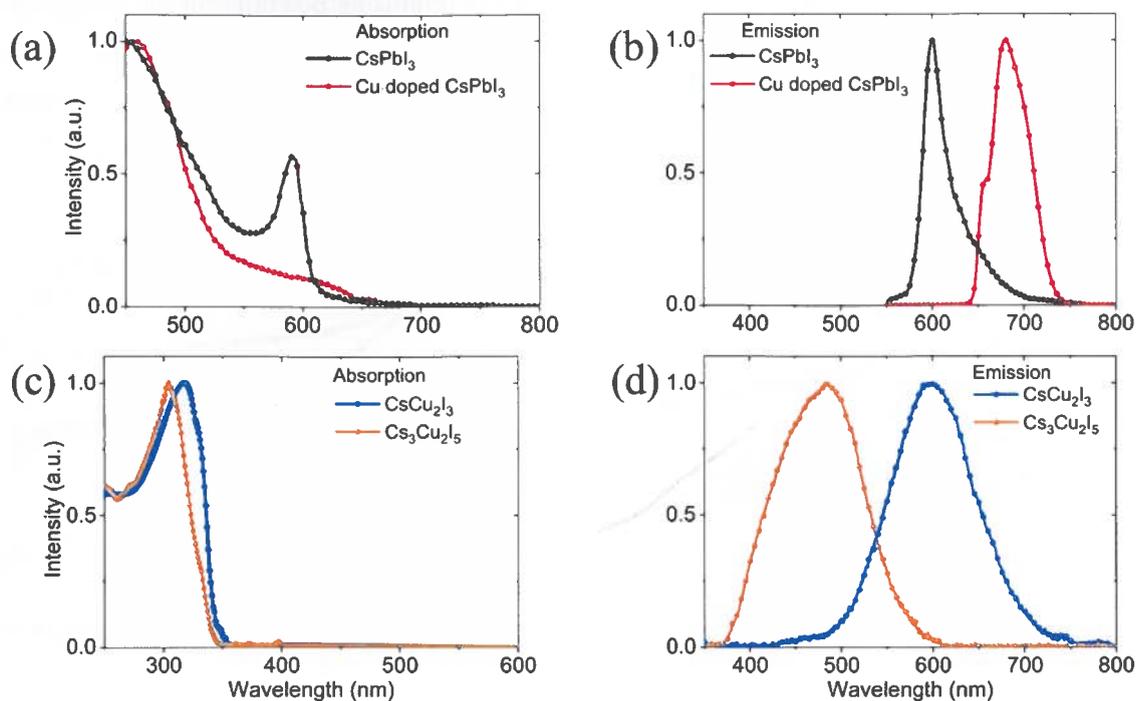

**Figure 3.** Absorption and photoluminescence spectra of the (a,b) CsPbI$_3$ and Cu-doped CsPbI$_3$, and (c,d) CsCu$_2$I$_3$ and Cs$_3$Cu$_2$I$_5$ measured at RT.

Figure 4a illustrates the time-resolved photoluminescence (TR-PL) decay characteristics of CsPbI$_3$ (black) and Cu-doped CsPbI$_3$ (red). The decay profile of the Cu-doped CsPbI$_3$ sample (red) exhibited a markedly faster decay rate than that of the pristine CsPbI$_3$ sample (black). Figure 4b illustrates the TR-PL decay characteristics of CsCu$_2$I$_3$ (blue) and Cs$_3$Cu$_2$I$_5$ (orange), demonstrating that reducing the perovskite dimensionality is associated with a qualitative acceleration in terms of the decay processes. As illustrated in Table 2, the $\tau_1$ value of the pristine CsPbI$_3$ is 0.8 ns, with a $\tau_1$ contribution of 45%. The effect of Cu doping resulted in a reduction of $\tau_1$ to 0.6 ns and an increase in the photon decay contribution to 53%. Additionally, the shortening of $\tau_2$ from 4.3 ns (contribution of 38%) in CsPbI$_3$ to 3.4 ns (contribution of 13%) after doping with Cu is demonstrated.

To obtain further insight, we present the impact of doping 5% Cu on the PLQY value, employing a well-established formula (see Equation S1) from prior research.[17] In an earlier study, the substitution of Pb atoms with 5.6% Cu resulted in a notable enhancement in PLQY, rising from 71 to 81%.[10] This was accompanied by a considerable increase in the radiative carrier decay lifetime. From this report shows that a reduction of 0.6% Cu resulted in no change in PLQY. These outcomes indicate that such dopant engineering led to a promising potential for the utilization of all-inorganic-based scintillators. Additionally, we compare the effect of Cu-doped to reduction dimension from previous report. Here we note that 1D of $CsCu_2I_3$ shows a significant reduction of PLQY to 13.59 ± 2.85 %, then elevated to the value of 47.19 ± 8.83 % for 0D of $Cs_2Cu_3I_5$.[17]

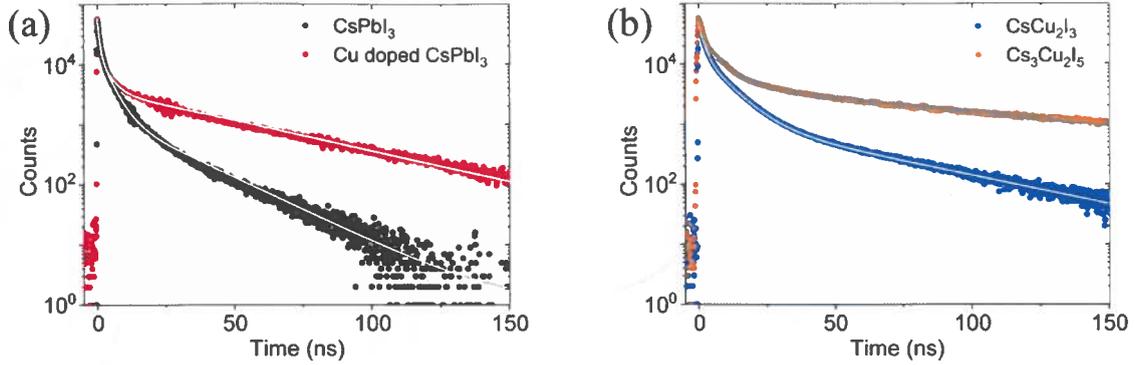

**Figure 4.** Time-resolved PL decay characteristics of (a) $CsPbI_3$ and Cu-doped $CsPbI_3$ (b) $CsCu_2I_3$ and $Cs_3Cu_2I_5$.

**Table 2.** Time-resolved PL decay characteristics of $CsPbI_3$, Cu-doped $CsPbI_3$, $CsCu_2I_3$, and $Cs_3Cu_2I_5$ where $\tau_1$ is the decay time, (%) is the contribution of the decay time, and $\tau_{avg}$ is the average decay time.

| Compounds | Fast ($\tau_1$ (%) ns) | Mid ($\tau_2$ (%) ns) | Slow ($\tau_3$ (%) ns) | $\tau_{avg}$ (ns) |
|---|---|---|---|---|
| $CsPbI_3$ | 0.8 ± 0.024 (45%) | 4.3 ± 0.13 (38%) | 20.0 ± 0.6 (17%) | 5.4 ± 0.16 |
| 5%Cu-$CsPbI_3$ | 0.6 ± 0.01 (53%) | 3.4 ± 0.1 (13%) | 42.9 ± 1.29 (34%) | 15.4 ± 0.46 |
| $CsCu_2I_3$ | 9.30 ± 0.01 (43%) | 54.50 ± 0.01 (30%) | 387.15 ± 6.05 (27%) | 124.88 ± 1.63 |
| $Cs_3Cu_2I_5$ | 10.90 ± 0.55 (3%) | 116.15 ± 0.35 (7%) | 924.85 ± 4.05 (90%) | 840.82 ± 3.65 |

To provide a point of comparison, we present the case of fully copper compounds with a dimensionality reduction depicted in $CsCu_2I_3$ and $Cs_3Cu_2I_5$. The dimensionality reduction is accompanied by an increase in the values of $\tau_1$ and $\tau_2$, from $9.30 \pm 0.01$ ns in 1D to $10.90 \pm 0.55$ ns in 0D and from $54.50 \pm 0.01$ ns in 1D to $116.15 \pm 0.35$ ns in 0D, respectively. We note that the contribution of $\tau_1$ and $\tau_2$ was diminished following a dimension reduction, due in part to the formation of vacancies within the crystalline structure. The elevated contribution $\tau_1$ in the Cu doping case substantiates the assertion that the partial substitution of lead in the B-site could compensate for the origin defect intrinsically governing the vacancy within the crystal structure. Although the average lifetime ($\tau_{avg}$) of $CsPbI_3$ is superior to that of low-dimensional samples or Cu doping, due to the substantial contribution to $\tau_3$, when compared with previous literature utilizing Ni doping, the $\tau_1$, $\tau_2$, and $\tau_3$ in Cu doping are shorter.[28] These findings illustrate that doping engineering is a considerably more efficacious approach for fast detection, both in terms of time and contribution, compared to reducing dimensionality through the substitution of Pb with Cu.

A series of temperature-dependent radioluminescence (RL) studies were conducted to examine the thermal influence on the emission characteristics of $CsPbI_3$, Cu-doped $CsPbI_3$, $CsCu_2I_3$, and $Cs_3Cu_2I_5$, as illustrated in Figures 5a-d, respectively (see also Figure S3). Upon X-ray excitation, the RL emission spectral lineshape of pristine $CsPbI_3$ highlighting two distinct wavelength regions: ~470-600 nm and ~690 nm (Figure 5a).[11] The first region, spanning the wavelengths between 470 and 600 nm exhibited an increase in intensity at temperatures between 10 and 200 K, followed by thermal quenching (TQ) at temperatures exceeding 250 K. The second region is associated with the RL emission center at ~668 nm, displaying an increase in intensity between 10-125 K, and a declining trend at the higher temperature (~200 K). The RL intensity exhibited a marked increase between the range of 225 K and 350 K. In addition to this increase in intensity, the second region of RL peak displays a blue shifted as the temperature is increased (Table S1). In the case of the Cu-doped $CsPbI_3$ (Figure 5b), the first region, ~470-600 nm, did not exhibit any notable intensity changes in comparison to the $CsPbI_3$. At the second region, the emission center at 700 nm exhibited an increase in intensity between temperatures of 10 K and ~350 K, accompanied by a blue shift by the increased temperature (Table S2). The $CsPbI_3$ tetragonal structure (β-phase) exhibited an RL peak at 716 nm.[12] A comparative analysis revealed that the tetragonal structure (β-phase) in Cu-doped CsPbI$_3$ exhibited a blue shift, as evidenced by the data presented.

The full-width at half-maximum (FWHM) analyses provide further interpretation on the TQ behavior as illustrated in Table. S1. Here we notice that the FWHM of CsPbI$_3$ exhibited a narrowing trend from 10 K to 200 K within the initial temperature range. The absence of peaks at 250 K indicates that the material had exhibited TQ at that temperature. The FWHM in the second region exhibited a narrowing from 10 K to 200 K, followed by a broadening from 250 K to 350 K. The concurrent narrowing and broadening of the FWHM during the temperature increase process in one region indicated a phase change, which in turn indicated a prominent phase instability in CsPbI$_3$ at the temperatures >200 K. The effect of doping demonstrated enhanced phase stability when there was no TQ from 10 K to 350 K. The FWHM of Cu-doped CsPbI$_3$ (Table S2) confirmed the stability emission with a stable FWHM value in the range of ~15-17 nm.

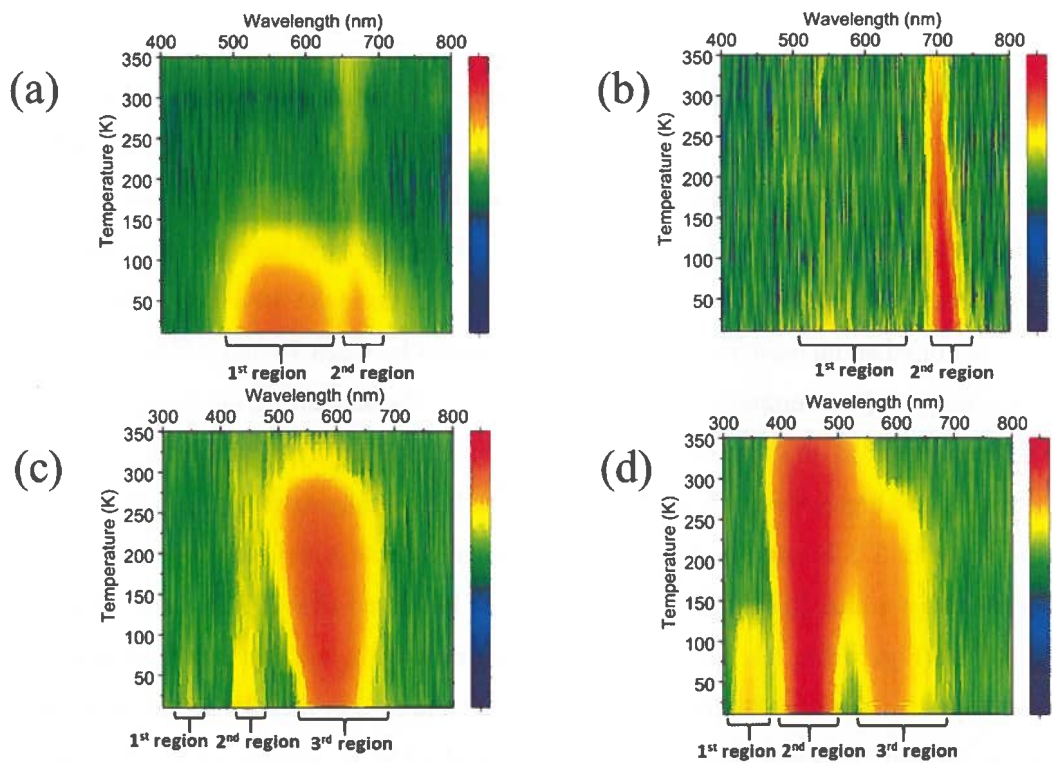

Figure 5. Temperature-dependent radioluminescence contour maps of (a) CsPbI$_3$ (b) Cu-doped CsPbI$_3$ (c) CsCu$_2$I$_3$ and, (d) Cs$_3$Cu$_2$I$_5$ covered the wavelengths of 350–600 nm and temperatures

of 10–350 K. The respective color bar correlates to the intensity of RL spectra, revealing a varied intensity in the under study of all-inorganic perovskite samples.

The impact of lowering the perovskite dimension on the scintillation properties of fully-converted copper-based perovskites, such as $CsCu_2I_3$ 1D and $Cs_2Cu_3I_5$ 0D are investigated. The results are illustrated in Figures. 5c and 5d, respectively. A tendency for a blue shift of RL emission peak and an increase in the TQ limit is observed. $CsCu_2I_3$ 1D and $Cs_3Cu_2I_5$ 0D exhibited three distinct emission regions, centered at the wavelengths of approximately 325 nm, 450 nm, and 575 nm, respectively. The initial region, which is centered at a wavelength of approximately 325 nm, exhibited a peak with relatively weak intensity in both 1D and 0D structure. The TQ of 1D is occurred at the lower temperature of 150 K, while the TQ of 0D occurred at 300 K. The second region, centered at ~450 nm, exhibited a peak intensity for 1D between 10 K and 300 K, with TQ occurring at temperatures above 300 K. In contrast, 0D displayed a peak intensity between 10 K and 350 K, accompanied by a broader luminescence and red shifting in the RL peak as the temperature increased. The third region, centered at ~575 nm, exhibited peak intensity between 10 K and 350 K for both 1D and 0D, with TQ occurring at temperatures above 350 K. However, 0D displayed a broader luminescence than 1D. The RL results on $CsPbI_3$ and CHPs yielded two and three emission regions, respectively. The presence of three emission regions in CHPs can be attributed to the adjacent RL and PL peaks, which suggests the possibility of a shared radiative recombination path.[17]

Table 3 presents a comparative analysis of the LY for $CsPbI_3$, Cu-doped $CsPbI_3$, $CsCu_2I_3$, and $Cs_3Cu_2I_5$. $CsPbI_3$ exhibited a LY value of 1.1 ± 0.5 photons per keV. The incorporation of 5% copper into the perovskite material led to a notable LY enhancement, reaching a value of 3.0 ± 0.8 photons/keV, which represents an increase of up to 172% compared to the initial measurement. The impact of Cu doping on LY performance has not been widely documented. However, when compared to the effect of Tl and In doping on $Cs_3Cu_2I_5$, their LY increased by 45.71% and 51.43%, respectively.[29] These results confirm that Cu doping is a promising solution to improve the performance of high-LY scintillators as previously suggested in the literature.[30] Although, reducing the dimensions resulted in a greater increased LY, reaching 12.5 ± 4.0 (1D) and 16.5 ± 3.5 photon/keV (0D).

Table 3. LY comparison between $CsPbI_3$, Cu-doped $CsPbI_3$, $CsCu_2I_3$, and $Cs_3Cu_2I_5$.

| Compounds | Light yields (photons/keV) | Ref. |
|---|---|---|
| $CsPbI_3$ | 1.9 ± 0.5 | 11 |
| 5%Cu-$CsPbI_3$ | 3.0 ± 0.8 | - |
| $CsCu_2I_3$ | 12.5 ± 4.0 | 17 |
| $Cs_3Cu_2I_5$ | 16.5 ± 3.5 | 17 |

**Conclusion**

In conclusion, the optical and scintillation properties of $CsPbI_3$, Cu-doped $CsPbI_3$, $CsCu_2I_3$, and $Cs_3Cu_2I_5$ have been investigated. We find that the structural impact on copper doping in the case of 3D all-inorganic lead-based perovskites resulted in a diminished absorption length compared to the observed dimensional reduction. The decay time characteristics of Cu doping exhibit a good promise of fast detection ($t_1$ 0.6 ns) attained according to the TRPL decay curves. To some extent, this achieves an improvement with respect to the lower dimension counterparts. Although the $\tau_{avg}$ of pristine $CsPbI_3$ is faster than Cu-doped $CsPbI_3$, here we need to emphasize that the emission behavior of Cu-doped has shown a superior thermal independence up to 50-350 K, in contrast to the case of pristine counterparts. In addition, emission tunable behavior is successfully increasing the LY value of $CsPbI_3$ to 3.0 ± 0.8 photons/keV, as well as the effect of reducing the perovskite dimensionality :$CsCu_2I_3$ 1D and $Cs_3Cu_2I_5$ 0D led to 12.5 ± 4.0 and 16.5 ± 3.5 photons/keV, respectively. Our attempt in Cu doping represents a promising avenue for the bright future development of lead-free scintillating material to come.

**Supporting Information.**

Rietveld refinement $CsPbI_3$ and Cu doped $CsPbI_3$; Radioluminescence spectrum $CsPbI_3$ and Cu doped $CsPbI_3$; Peak position and FWHM value of $CsPbI_3$ and Cu doped $CsPbI_3$.

**Accession Codes**

CCDC 2304835 and 2303801 contain the supplementary crystallographic data for this paper. These data can be obtained free of charge via www.ccdc.cam.ac.uk/data_request/cif, by emailing at data_request@ccdc.cam.ac.uk, or by contacting the Cambridge Crystallographic Data Centre, 12 Union Road, Cambridge CB2 1EZ, UK; fax: + 44 1223 336033.

**AUTHOR INFORMATION**

Corresponding Authors

Somnath Mahato − Lukasiewicz Research Network - PORT Polish Center for Technology Development, Stabłowicka 147, Wroclaw 54-066, Poland; orcid.org/0000-0002-7208-283X

Email: somnath.mahato@port.lukasiewicz.gov.pl

Arramel - Center of Excellence Applied Physics and Chemistry, Nano Center Indonesia, South Tangerang 15314, Indonesia; orcid.org/0000-0003-4125-6099; Email: *arramel@nano.or.id*

Muhammad Danang Birowosuto − *Łukasiewicz Research Network − PORT Polish Center for Technology Development, Wrocław 54-066, Poland*; orcid.org/0000- 0002-9997-6841; Email: muhammad.birowosuto@port.lukasiewicz.gov.pl

Authors

David Hadid Sidiq − Center of Excellence Applied Physics and Chemistry, Nano Center Indonesia, South Tangerang 15314, Indonesia; orcid.org/0009-0005-2824-7165

Tobias Haposan − Department of Materials Science and Engineering, National University of Singapore, 9 Engineering Drive 1, Singapore 117575, Singapore; orcid.org/0000-0002-1907-6609

Michal Makowski − Lukasiewicz Research Network - PORT Polish Center for Technology Development, Stabłowicka 147, Wroclaw 54-066, Poland; orcid.org/0000-0001-7758-2326

Dominik Kowal − Lukasiewicz Research Network - PORT Polish Center for Technology Development, Stabłowicka 147, Wroclaw 54-066, Poland; orcid.org/0000-0001-9424-0416

Marcin Eugeniusz Witkowski − Institute of Physics, Faculty of Physics, Astronomy, and Informatics, Nicolaus Copernicus University in Torun, ul. Grudziadzka 5, 87-100 Torun, Poland; orcid.org/0000-0001-6282-8610

Winicjusz Drozdowski − Institute of Physics, Faculty of Physics, Astronomy, and Informatics, Nicolaus Copernicus University in Torun, ul. Grudziadzka 5, 87-100 Torun, Poland; orcid.org/0000-0002-6207-4801

Author Contributions

Conceptualisation: A., and M.D.B.; data curation: S. M., M. M., D. K., and M. E. W; formal analysis: D. H. S., T. H. and A.; investigation: D. H. S., T. H., and A.; methodology: W. D., and M. D. B.; project administration: A.; resources: M. D. B., and W. D.; supervision: M.D.B. and A.; visualisation: D. H. S,. and T.H.; writing—original draft: D. H. S. and A.; writing—review and editing: M. D. B. and A. All authors have read and agreed to the published version of the manuscript.

**Notes**

The authors declare no competing financial interest.

**ACKNOWLEDGMENTS**

S.M. acknowledges the support by the European Commission (Agreement No. 945339) and the Polish National Science Centre under the Marie Skłodowska-Curie COFUND grant (POLONEZ BIS 2) under Agreement No: UMO-2022/45/P/ST3/04170. M.D.B. acknowledge funding from National Science Center, Poland, under Grant OPUS-24 No. 2022/47/B/ST5/01966. S.M. acknowledges to Baidyanath Roy at Indian Institute of Technology Kharagpur (India) for synthesis of $CsPbI_3$ nanocrystals and Cu doping samples. Author from Nano Center Indonesia express their gratitude to PT Nanotech Indonesia Global Tbk for the start-up research grant.

## ABBREVIATIONS

TEM, Transmission electron microscopy; HRTEM, high-resolution TEM; PL, photoluminescence; TRPL, Time-resolved photoluminescence; RL, Radioluminescence; CHPs, Copper halide perovskites; TQ, Thermal quenching

## REFERENCES

(1) Mykhaylyk, V. B.; Rudko, M.; Kraus, H.; Kapustianyk, V.; Kolomiets, V.; Vitoratou, N.; Chornodolskyy, Y.; Voloshinovskii, A. S.; Vasylechko, L. Ultra-Fast Low Temperature Scintillation and X-Ray Luminescence of CsPbCl$_3$ Crystals. *J. Mater. Chem. C* **2023**, *11* (2), 656–665. https://doi.org/10.1039/D2TC04631H.

(2) Protesescu, L.; Yakunin, S.; Bodnarchuk, M. I.; Krieg, F.; Caputo, R.; Hendon, C. H.; Yang, R. X.; Walsh, A.; Kovalenko, M. V. Nanocrystals of Cesium Lead Halide Perovskites (CsPbX$_3$, X = Cl, Br, and I): Novel Optoelectronic Materials Showing Bright Emission with Wide Color Gamut. *Nano Lett.* **2015**, *15* (6), 3692–3696. https://doi.org/10.1021/nl5048779.

(3) Bose, S.; Mahato, S.; Roy, B.; Singha, T.; Srivastava, S. K.; Ray, S. K. Unveiling Electron Dose-Induced Phase Decomposition and Energy Kinetics in Cu-Doped CsPbI$_3$ Nanocrystals. *ACS Appl. Nano Mater.* **2024**, *7* (6), 6020–6028. https://doi.org/10.1021/acsanm.3c05929.

(4) Sebastian, M.; Peters, J. A.; Stoumpos, C. C.; Im, J.; Kostina, S. S.; Liu, Z.; Kanatzidis, M. G.; Freeman, A. J.; Wessels, B. W. Excitonic Emissions and Above-Band-Gap Luminescence in the Single-Crystal Perovskite Semiconductors CsPbBr$_3$ and CsPbCl$_3$. *Phys. Rev. B* **2015**, *92* (23), 235210. https://doi.org/10.1103/PhysRevB.92.235210.

(5) Birowosuto, M. D.; Cortecchia, D.; Drozdowski, W.; Brylew, K.; Lachmanski, W.; Bruno, A.; Soci, C. X-Ray Scintillation in Lead Halide Perovskite Crystals. *Sci. Rep.* **2016**, *6* (1), 37254. https://doi.org/10.1038/srep37254.

(6) Birowosuto, M. D.; Maddalena, F.; Aozhen, X.; Witkowski, M.; Makowski, M.; Drozdowski, W.; Coquet, P.; Dujardin, C.; Dang, C. Scintillators from Solution-Processable Perovskite Halide Single Crystals or Quantum Dots: The Good, the Bad, and the Ugly. In *Hard X-Ray, Gamma-Ray, and Neutron Detector Physics XXII*; Fiederle, M., Burger, A., Payne, S. A., Eds.; SPIE: Online Only, United States, 2020; p 39. https://doi.org/10.1117/12.2569663.

(7) Kobayashi, M.; Omata, K.; Sugimoto, S.; Tamagawa, Y.; Kuroiwa, T.; Asada, H.; Takeuchi, H.; Kondo, S. Scintillation Characteristics of CsPbCl$_3$ Single Crystals. *Nucl. Instrum. Methods Phys. Res. Sect. Accel. Spectrometers Detect. Assoc. Equip.* **2008**, *592* (3), 369–373. https://doi.org/10.1016/j.nima.2008.04.079.

(8) Chen, Z.; Zhou, B.; Yuan, J.; Tang, N.; Lian, L.; Qin, L.; Zhu, L.; Zhang, J.; Chen, R.; Zang, J. Cu$^{2+}$-Doped CsPbI$_3$ Nanocrystals with Enhanced Stability for Light-Emitting Diodes. *J. Phys. Chem. Lett.* **2021**, *12* (12), 3038–3045. https://doi.org/10.1021/acs.jpclett.1c00515.

(9) Diroll, B. T.; Zhou, H.; Schaller, R. D. Low-Temperature Absorption, Photoluminescence, and Lifetime of CsPbX$_3$ (X = Cl, Br, I) Nanocrystals. *Adv. Funct. Mater.* **2018**, *28* (30), 1800945. https://doi.org/10.1002/adfm.201800945.

(10) Roy, B.; Mahato, S.; Bose, S.; Ghorai, A.; Srivastava, S. K.; Das, N. C.; Ray, S. K. Cu-Doping Induced Phase Transformation in CsPbI$_3$ Nanocrystals with Enhanced Structural Stability and Photoluminescence Quantum Yield. *Chem. Mater.* **2023**, *35* (4), 1601–1609. https://doi.org/10.1021/acs.chemmater.2c03095.

(11) Maddalena, F.; Xie, A.; Chin, X. Y.; Begum, R.; Witkowski, M. E.; Makowski, M.; Mahler, B.; Drozdowski, W.; Springham, S. V.; Rawat, R. S.; Mathews, N.; Dujardin, C.; Birowosuto, M. D.; Dang, C. Deterministic Light Yield, Fast Scintillation, and Microcolumn Structures in Lead Halide Perovskite Nanocrystals. *J. Phys. Chem. C* **2021**, *125* (25), 14082–14088. https://doi.org/10.1021/acs.jpcc.1c03392.

(12) Chen, D.; Huang, D.; Yang, M.; Xu, K.; Hu, J.; Xu, F.; Liang, S.; Zhu, H. Room-Temperature Direct Synthesis of Tetragonal β-CsPbI$_3$ Nanocrystals. *Adv. Opt. Mater.* **2022**, *10* (2), 2101869. https://doi.org/10.1002/adom.202101869.

(13) Yao, J.-S.; Ge, J.; Wang, K.-H.; Zhang, G.; Zhu, B.-S.; Chen, C.; Zhang, Q.; Luo, Y.; Yu, S.-H.; Yao, H.-B. Few-Nanometer-Sized α-CsPbI$_3$ Quantum Dots Enabled by Strontium Substitution and Iodide Passivation for Efficient Red-Light Emitting Diodes. *J. Am. Chem. Soc.* **2019**, *141* (5), 2069–2079. https://doi.org/10.1021/jacs.8b11447.

(14) Roy, S.; Prasad, E. Effect of Co$^{2+}$ Doping on Optical Property and Exciton–Phonon Coupling in CsPbI$_3$ Perovskite Nanocrystals. *J. Phys. Chem. C* **2023**, *127* (42), 20802–20810. https://doi.org/10.1021/acs.jpcc.3c05555.

(15) Shen, X.; Zhang, Y.; Kershaw, S. V.; Li, T.; Wang, C.; Zhang, X.; Wang, W.; Li, D.; Wang, Y.; Lu, M.; Zhang, L.; Sun, C.; Zhao, D.; Qin, G.; Bai, X.; Yu, W. W.; Rogach, A. L. Zn-Alloyed CsPbI$_3$ Nanocrystals for Highly Efficient Perovskite Light-Emitting Devices. *Nano Lett.* **2019**, *19* (3), 1552–1559. https://doi.org/10.1021/acs.nanolett.8b04339.

(16) Wang, B.; Novendra, N.; Navrotsky, A. Energetics, Structures, and Phase Transitions of Cubic and Orthorhombic Cesium Lead Iodide (CsPbI$_3$) Polymorphs. *J. Am. Chem. Soc.* **2019**, *141* (37), 14501–14504. https://doi.org/10.1021/jacs.9b05924.

(17) Haposan, T.; Arramel, A.; Maulida, P. Y. D.; Hartati, S.; Afkauni, A. A.; Mahyuddin, M. H.; Zhang, L.; Kowal, D.; Witkowski, M. E.; Drozdowski, K. J.; Makowski, M.; Drozdowski, W.; Diguna, L. J.; Birowosuto, M. D. All-Inorganic Copper-Halide Perovskites for Large-Stokes Shift and Ten-Nanosecond-Emission Scintillators. *J. Mater. Chem. C* **2024**, *12* (7), 2398–2409. https://doi.org/10.1039/D3TC03977C.

(18) Momma, K.; Izumi, F. *VESTA 3* for Three-Dimensional Visualization of Crystal, Volumetric and Morphology Data. *J. Appl. Crystallogr.* **2011**, *44* (6), 1272–1276. https://doi.org/10.1107/S0021889811038970.

(19) Fang, Z.; Shang, M.; Hou, X.; Zheng, Y.; Du, Z.; Yang, Z.; Chou, K.-C.; Yang, W.; Wang, Z. L.; Yang, Y. Bandgap Alignment of α-CsPbI$_3$ Perovskites with Synergistically Enhanced Stability and Optical Performance via B-Site Minor Doping. *Nano Energy* **2019**, *61*, 389–396. https://doi.org/10.1016/j.nanoen.2019.04.084.

(20) Maddalena, F.; Xie, A.; Arramel; Witkowski, M. E.; Makowski, M.; Mahler, B.; Drozdowski, W.; Mariyappan, T.; Springham, S. V.; Coquet, P.; Dujardin, C.; Birowosuto, M. D.; Dang, C. Effect of Commensurate Lithium Doping on the Scintillation of Two-Dimensional Perovskite Crystals. *J. Mater. Chem. C* **2021**, *9* (7), 2504–2512. https://doi.org/10.1039/D0TC05647B.

(21) Guo, N.; Liu, L.; Cao, G.; Xing, S.; Liang, J.; Chen, J.; Tan, Z.; Shang, Y.; Lei, H. Enhancing Emission and Stability in Na-Doped Cs$_3$Cu$_2$I$_5$ Nanocrystals. *Nanomaterials* **2024**, *14* (13), 1118. https://doi.org/10.3390/nano14131118.

(22) Wen, X.; Gao, Q.; Wang, Q.; Chewpraditkul, W.; Korjik, M.; Kurosawa, S.; Buryi, M.; Babin, V.; Wu, Y. Highly Efficient In(I) Doped Cs$_3$Cu$_2$I$_5$ Single Crystals for Light-Emitting Diodes and Gamma Spectroscopy Applications. *Opt. Mater. X* **2024**, *23*, 100335. https://doi.org/10.1016/j.omx.2024.100335.

(23) Lin, R.; Zhu, Q.; Guo, Q.; Zhu, Y.; Zheng, W.; Huang, F. Dual Self-Trapped Exciton Emission with Ultrahigh Photoluminescence Quantum Yield in $CsCu_2I_3$ and $Cs_3Cu_2I_5$ Perovskite Single Crystals. *J. Phys. Chem. C* **2020**, *124* (37), 20469–20476. https://doi.org/10.1021/acs.jpcc.0c07435.

(24) Qiu, F.; Lei, Y.; Jin, Z. Copper-Based Metal Halides for X-Ray and Photodetection. *Front. Optoelectron.* **2022**, *15* (1), 47. https://doi.org/10.1007/s12200-022-00048-x.

(25) Gu, Y.; Yao, X.; Geng, H.; Guan, G.; Hu, M.; Han, M. Highly Transparent, Dual-Color Emission, Heterophase $Cs_3Cu_2I_5$/$CsCu_2I_3$ Nanolayer for Transparent Luminescent Solar Concentrators. *ACS Appl. Mater. Interfaces* **2021**, *13* (34), 40798–40805. https://doi.org/10.1021/acsami.1c07686.

(26) Zhang, X.; Zhou, B.; Chen, X.; Yu, W. W. Reversible Transformation between $Cs_3Cu_2I_5$ and $CsCu_2I_3$ Perovskite Derivatives and Its Anticounterfeiting Application. *Inorg. Chem.* **2022**, *61* (1), 399–405. https://doi.org/10.1021/acs.inorgchem.1c03021.

(27) Kowal, D.; Makowski, M.; Witkowski, M. E.; Cala', R.; Kuddus Sheikh, M. A.; Mahyuddin, M. H.; Auffray, E.; Drozdowski, W.; Cortecchia, D.; Birowosuto, M. D. $PEA_2PbI_4$: Fast Two-Dimensional Lead Iodide Perovskite Scintillator with Green and Red Emission. *Mater Today Chem* **2023**, *29*, 101455. https://doi.org/10.1016/j.mtchem.2023.101455.

(28) Liu, M.; Jiang, N.; Huang, H.; Lin, J.; Huang, F.; Zheng, Y.; Chen, D. $Ni^{2+}$-Doped $CsPbI_3$ Perovskite Nanocrystals with near-Unity Photoluminescence Quantum Yield and Superior Structure Stability for Red Light-Emitting Devices. *Chem. Eng. J.* **2021**, *413*, 127547. https://doi.org/10.1016/j.cej.2020.127547.

(29) Qu, J.; Xu, S.; Shao, H.; Xia, P.; Lu, C.; Wang, C.; Ban, D. Recent Progress of Copper Halide Perovskites: Properties, Synthesis and Applications. *J. Mater. Chem. C* **2023**, *11* (19), 6260–6275. https://doi.org/10.1039/D3TC00503H.

(30) Gu, R.; Han, K.; Jin, J.; Zhang, H.; Xia, Z. Surfactant-Assisted Synthesis of Hybrid Copper(I) Halide Nanocrystals for X-Ray Scintillation Imaging. *Chem. Mater.* **2024**, *36* (6), 2963–2970. https://doi.org/10.1021/acs.chemmater.4c00020.

(31) Amgarou, K.; Herranz, M. State-of-the-Art and Challenges of Non-Destructive Techniques for in-Situ Radiological Characterization of Nuclear Facilities to Be Dismantled. *Nucl. Eng. Technol.* **2021**, *53* (11), 3491–3504. https://doi.org/10.1016/j.net.2021.05.031.

(32) Boudjemila L.; Aleshin A. N.; Malyshkin V. M.; Aleshin P. A.; Shcherbakov I. P.; Petrov V. N.; Terukov E. I. Electrical and Optical Characteristics of $CsPbI_3$ and $CsPbBr_3$ Lead Halide Perovskite Nanocrystal Films Deposited on c-Si Solar Cells for Photovoltaic Applications. *Phys. Solid State* **2022**, *64* (11), 1670. https://doi.org/10.21883/PSS.2022.11.54189.418.

(33) Adl, H. P.; Gorji, S.; Habil, M. K.; Suárez, I.; Chirvony, V. S.; Gualdrón-Reyes, A. F.; Mora-Seró, I.; Valencia, L. M.; De La Mata, M.; Hernández-Saz, J.; Molina, S. I.; Zapata-Rodríguez, C. J.; Martínez-Pastor, J. P. Purcell Enhancement and Wavelength Shift of Emitted Light by $CsPbI_3$ Perovskite Nanocrystals Coupled to Hyperbolic Metamaterials. *ACS Photonics* **2020**, *7* (11), 3152–3160. https://doi.org/10.1021/acsphotonics.0c01219.

(34) Maddalena, F.; Tjahjana, L.; Xie, A.; Arramel; Zeng, S.; Wang, H.; Coquet, P.; Drozdowski, W.; Dujardin, C.; Dang, C.; Birowosuto, M. Inorganic, Organic, and Perovskite Halides with Nanotechnology for High–Light Yield X- and γ-Ray Scintillators. *Crystals* **2019**, *9* (2), 88. https://doi.org/10.3390/cryst9020088.

(35) Moseley, O. D. I.; Doherty, T. A. S.; Parmee, R.; Anaya, M.; Stranks, S. D. Halide Perovskites Scintillators: Unique Promise and Current Limitations. *J. Mater. Chem. C* **2021**, *9* (35), 11588–11604. https://doi.org/10.1039/D1TC01595H.

(36) Skurlov, I. D.; Sokolova, A. V.; Tatarinov, D. A.; Parfenov, P. S.; Kurshanov, D. A.; Ismagilov, A. O.; Koroleva, A. V.; Danilov, D. V.; Zhizhin, E. V.; Mikushev, S. V.; Tcypkin, A. N.; Fedorov, A. V.; Litvin, A. P. Engineering the Optical Properties of CsPbBr3 Nanoplatelets through $Cd^{2+}$ Doping. *Materials* **2022**, *15* (21), 7676. https://doi.org/10.3390/ma15217676.

(37) Livakas, N.; Toso, S.; Ivanov, Y. P.; Das, T.; Chakraborty, S.; Divitini, G.; Manna, L. $CsPbCl_3 \rightarrow CsPbI_3$ Exchange in Perovskite Nanocrystals Proceeds through a Jump-the-Gap Reaction Mechanism. *J. Am. Chem. Soc.* **2023**, *145* (37), 20442–20450. https://doi.org/10.1021/jacs.3c06214.

(38) Wibowo, A.; Sheikh, M. A. K.; Diguna, L. J.; Ananda, M. B.; Marsudi, M. A.; Arramel, A.; Zeng, S.; Wong, L. J.; Birowosuto, M. D. Development and Challenges in Perovskite Scintillators for High-Resolution Imaging and Timing Applications. *Commun. Mater.* **2023**, *4* (1), 21. https://doi.org/10.1038/s43246-023-00348-5.

(39) Xu, K.; Lin, C. C.; Xie, X.; Meijerink, A. Efficient and Stable Luminescence from $Mn^{2+}$ in Core and Core–Isocrystalline Shell $CsPbCl_3$ Perovskite Nanocrystals. *Chem. Mater.* **2017**, *29* (10), 4265–4272. https://doi.org/10.1021/acs.chemmater.7b00345.

(40) Zhao, X.; Tang, T.; Xie, Q.; Gao, L.; Lu, L.; Tang, Y. First-Principles Study on the Electronic and Optical Properties of the Orthorhombic $CsPbBr_3$ and $CsPbI_3$ with *Cmcm* Space Group. *New J. Chem.* **2021**, *45* (35), 15857–15862. https://doi.org/10.1039/D1NJ02216D.

(41) Zou, S.; Liu, Y.; Li, J.; Liu, C.; Feng, R.; Jiang, F.; Li, Y.; Song, J.; Zeng, H.; Hong, M.; Chen, X. Stabilizing Cesium Lead Halide Perovskite Lattice through Mn(II) Substitution for Air-Stable Light-Emitting Diodes. *J. Am. Chem. Soc.* **2017**, *139* (33), 11443–11450. https://doi.org/10.1021/jacs.7b04000.

(42) Ye, W.; Yong, Z.; Go, M.; Kowal, D.; Maddalena, F.; Tjahjana, L.; Wang, H.; Arramel, A.; Dujardin, C.; Birowosuto, M. D.; Wong, L. J. The Nanoplasmonic Purcell Effect in Ultrafast and High-Light-Yield Perovskite Scintillators. *Adv. Mater.* **2024**, *36* (25), 2309410. https://doi.org/10.1002/adma.202309410.

For Table of Contents Only

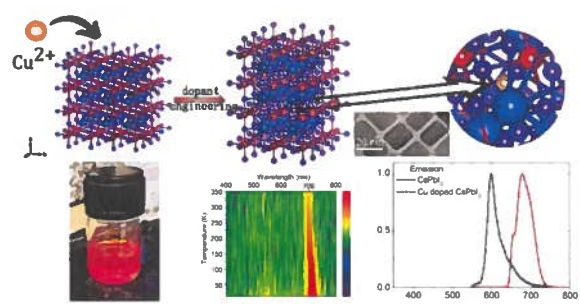

22